\documentclass[final,transmag,twoside]{IEEEtran}
\usepackage{cite}

\ifCLASSINFOpdf
\else
\fi
\usepackage[cmex10]{amsmath}
\usepackage[usenames]{color}
\usepackage{verbatim}
\usepackage{amsthm,amssymb}
\usepackage{psfrag}
\usepackage{bm}

\newcommand{\comp}[2]{#1$_{1-x}$#2$_x$MnO$_3$}
\newcommand{\Com}[4]{#1$_{#2}$#3$_{#4}$MnO$_3$}
\newcommand{\LCom}[4]{#1$_{#2}$#3$_{#4}$MnO$_4$}
\newcommand{\BCom}[5]{#1(#2$_{#3}$#4$_{#5}$)$_2$Mn$_2$O$_7$}
\newcommand{\Mn}[1]{Mn$^{#1+}$}
\newcommand{\Sn}{\ensuremath{\boldsymbol{S}}}
\newcommand{\Fp}{\ensuremath{\textsf{F$^{\,\prime}$}}}
\newcommand{\F}{\ensuremath{\textsf{F}}}
\newcommand{\A}{\ensuremath{\textsf{A}}}
\newcommand{\Ad}{\ensuremath{\textsf{A$_{\textsf{d}}$}}}
\newcommand{\K}{\ensuremath{\textsf{K}}}
\newcommand{\D}{\ensuremath{\textsf{D}}}

\begin{document}
\title{Quantum magnons of the intermediate phase of half-doped manganite oxides}

\author{\IEEEauthorblockN{
Ivon R. Buitrago\IEEEauthorrefmark{1,2},
Cecilia I. Ventura\IEEEauthorrefmark{1,3}, and
Luis O. Manuel\IEEEauthorrefmark{4}}

\IEEEauthorblockA{\IEEEauthorrefmark{1} (CONICET) Centro At\'omico Bariloche, CNEA, Argentina.}

\IEEEauthorblockA{\IEEEauthorrefmark{2} Instituto Balseiro, Univ. Nac. de Cuyo and CNEA, Argentina}

\IEEEauthorblockA{\IEEEauthorrefmark{3} Univ. Nac. de R\'{\i}o Negro, Bariloche, Argentina}
\IEEEauthorblockA{\IEEEauthorrefmark{4}  Instituto de F\'{\i}sica Rosario (CONICET-UNR), Rosario, Argentina.} 

\thanks{Corresponding author: I. Buitrago (email: ivonnebuitrago@cab.cnea.gov.ar)}}

%



\IEEEtitleabstractindextext{
\begin{abstract}
At half doping, the ground state of three-dimensional manganite perovskite oxides like \comp{R}{Ca}, where R is a trivalent ion such as La, Pr, etc, is still unclear. Many experimental findings agree better with the combined magnetic, charge, and orbital order characteristic of the ``intermediate phase", introduced by Efremov et al. in 2004 [Nature Mats. 3, 853]. This phase consists of spin dimers (thus incorporating aspects of the Zener polaron phase (ZP) proposed in 2002 by Daoud-Aladine et al. [Phys. Rev. Lett. 89, 097205]), though formed by a pair of parallel Mn spins of different magnitude, in principle (thereby allowing for a degree of Mn charge disproportionation: not necessarily as large as that of \Mn{3}-\Mn{4} in Goodenough's original CE phase [Phys. Rev. 100, 564 (1955)]). In the intermediate phase, consecutive spin dimers localed along the planar zig-zag chains are oriented at a constant relative angle $\phi$ between them. Varying Mn-charge disproportionation and $\phi$, the intermediate phase 
should allow to continuously interpolate between the two limiting cases of the CE phase and the dimer phase denoted as ``orthogonal intermediate $\pi/2-$phase".
It is not easy to find a microscopic model able to describe the phenomenological intermediate phase adequately for the spin, charge, and orbital degrees of freedom simultaneously. Here, we study the quantum spin excitations of a planar model of interacting localized spins, which we found can stabilize the intermediate phase classically. We compare the quantum magnons of the intermediate phase with those of the CE and orthogonal $\pi/2$ phases, in the context of recent experimental results.
\end{abstract}

}

\maketitle
\IEEEdisplaynontitleabstractindextext

%
\IEEEpeerreviewmaketitle

\section{Introduction}
\IEEEPARstart{A}{t half-doping}, three-dimensional perovskite manganites present difficulties for magnetic excitation measurements, such as inelastic neutron scattering (INS), which allow to indirectly probe the magnetic ground state too. For laminar half-doped manganites, experimental INS progress was achieved in 2006 by Senff et al.~\cite{Senff}, who were able to measure well defined magnons in \LCom{La}{0.5}{Sr}{1.5} below 40 meV, and compared various possible fits~\cite{Senff}. In 2011, INS low energy data from \Com{Nd}{0.5}{Sr}{0.5} single crystals were obtained by Ulbrich et al.~\cite{Ulbrich}, and, very recently \mbox{Johnstone} et al.~\cite{Johnstone} presented INS data up to 100 meV, including higher energy branches for the half-doped bilayer man\-ga\-nite \BCom{Pr}{Ca}{0.9}{Sr}{0.1}. These works show evidence of a charge, orbital, and spin ordering described by a CE-like phase, as proposed in 1955 by Goodenough~\cite{Goodenough} but with a markedly reduced charge disproportionation. Goodenough's CE 
phase is characterized by planes with a checkerboard charge order for \Mn{3} and \Mn{4} ions, while between consecutive planes equal charges are stacked one on top of each other. This leads to a magnetic arrangement of spins in ferromagnetic (FM) zig-zag chains in the planes, antiferromagnetically (AF) coupled between them. On the other hand, the neutron diffraction study of \Com{Pr}{0.6}{Ca}{0.4} by Daoud-Aladine et al.~\cite{Aladine} in 2002, did not find evidence for CE ordering. Instead, to explain their measurements, in Ref.~\cite{Aladine} the Zener polaron phase (ZP) was proposed, in which a delocalized electron is shared between each pair of Mn neighbour sites along a zig-zag chain, through the Zener double-exchange mechanism, so that dimers are formed with effective intermediate valence charge \Mn{3.5} present at each site. In previous work, we compared the magnons for charge-ordered and dimer phases for 3D~\cite{Ventura} and 2D half-doped manganites~\cite{Buitrago}, whereas the ZP phase has been 
found theoretically in Hartree-Fock electronic structure calculations~\cite{Patterson}. 

Since both the CE and ZP models are in agreement with measured results for
different systems, in 2004 Efremov et al.~\cite{Efremov} proposed the
intermediate phase, as a superposition of both, which besides attempting to
reconcile aspects of the CE and dimer phases, might explain various puzzling
experimental results in manganites~\cite{Mercone,Hill}. The intermediate
phase consists of spin dimers along the planar zig-zag chains, formed by pairs
of parallel Mn spins of different magnitude in principle. Consecutive spin
dimers in a chain are oriented at a constant relative angle $\phi$. Varying
 the Mn-charge disproportionation and angle $\phi$, the intermediate phase should allow to continuously interpolate between two limiting cases: Goodenough's CE phase ($\phi=0$) and the dimer phase denoted as the ``orthogonal intermediate $\pi/2$-phase" ($\phi=\pi/2$). Notice that in the latter $\pi/2$-phase~\cite{Efremov} consecutive dimers would have their spins oriented perpendicularly, in contrast to the ZP phase~\cite{Aladine} where all dimers have parallel spins: these two dimer phases as well as the CE phase were analyzed as possible electronic ground states for the fits to the magnons measured recently~\cite{Johnstone}.

With this scenario, we decided to calculate the quantum spin excitations of a
model based on interacting localized electronic spins, which we have found to
stabilize the intermediate phase in its classical limit, as discussed in next
section. In such model, we investigated which magnetic coupling parameters
would suffice to describe the intermediate phase, and allow a continuous
interpolation, between the CE and orthogonal intermediate phases, as Efremov
et al. intended~\cite{Efremov}. Ref.~\cite{Giovanetti} first focused on this
problem, though they found that phonons coupling to the electrons and
electronic spin delocalization, were needed to stabilize the intermediate phase at intermediate angles. 

\section{Model for the intermediate phase:\\classical analysis}
\begin{figure}[!h]
 \centering\scalebox{0.6}{
 \includegraphics{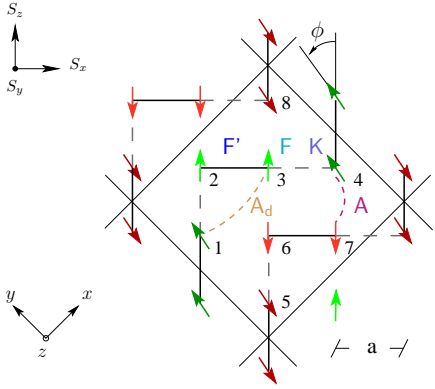}}
 \caption{(color online). Schematic representation of the intermediate phase. Magnetic couplings as described in the text. Spins of magnitude $S_1$: on the odd-numbered sites of the unit cell shown and of magnitude $S_2$: on the even-numbered sites.}
 \label{figure1}
\end{figure}

To describe the intermediate phase of Efremov et al.~\cite{Efremov} for laminar half-doped manganites, we analyzed a two-dimensional model of interacting localized spins, schematically shown in Fig~\ref{figure1}.

Our model consists of spin dimers along the zig-zag chains, formed by a pair
of nearest-neighbour (NN) Mn spins of different magnitude, e.g. $S_1=2$ (\Mn{3}) and $S_2=3/2$ (\Mn{4}) in the CE limit. Consecutive dimers in each chain are oriented at a constant relative angle, $\phi$, between them, shown in Fig.~\ref{figure1}, and the Mn-charge disproportionation is proportional to difference $\Delta=S_1-S_2=0.5\,\cos\phi$ between the localized spin magnitudes (e.g. $\Delta=0.5$ for the CE phase, whereas $\Delta=0$ for the orthogonal-$\pi/2$ and ZP phases). The magnetic couplings taken into account are: two ferromagnetic (FM) nearest-neighbour (NN) couplings, intra-dimer $\Fp$ and inter-dimer $\F$ along the zig-zag chains; one antiferromagnetic (AF) coupling $\A$ between chains; and single-ion anisotropy $\D$, as in our previous work for the CE phase~\cite{Buitrago} and the experiments in laminar \LCom{La}{0.5}{Sr}{1.5}~\cite{Senff}. In addition, we found that in order to stabilize the intermediate phase classically, as we will show next, we needed to include an AF next-nearest-neighbour 
(NNN) coupling, $\Ad$, between consecutive \Mn{3} ions within a chain, and a biquadratic inter-dimer coupling, $\K$, along chains. We found that the latter is the key to classically stabilize a perpendicular orientation between the spins of consecutive dimers along a chain, and it was previously included for perovskite manganites.~\cite{Cieplak}. 
The resulting Hamiltonian reads:
\begin{gather}
  \begin{split}
    H=-&\,\Fp\sum_{\langle i,j\rangle\in C,D}\,\Sn_i\cdot \Sn_j -\,\F\sum_{\langle i,j\rangle\in C,\notin\,D} \Sn_i\cdot \Sn_j^{\prime}\\[1mm]
    +&\,\A\sum_{\langle i,j\rangle/\notin C} \Sn_i\cdot \Sn_j^{\prime} +\,\Ad\sum_{\langle i,j\rangle/\in\,C} \Sn_i\cdot \Sn_j^{\prime} \\[1mm]
    +&\,\K\sum_{\langle i,j\rangle\in C,\notin\,D} \big(\Sn_i\cdot \Sn_j^{\prime}\big)^2
    -\,\D\sum_i \Sn_{iz}^2
  \end{split}
\label{EQ1}
\end{gather}
where $C$ means in one chain, $D$ in a dimer, $\Sn_j^{\prime}$ is used for rotated spins, and all non-zero coupling parameters are positive. 

To obtain the CE ($\phi=0$, $S_1=2$, $S_2=3/2$, $\F=\Fp$) and orthogonal intermediate ($\phi=\pi/2$, $S_1=S_2=7/4$, $\F=\Fp/2$) phases as limiting cases of the intermediate phase, in the following we use for the angle dependence of the spin magnitudes (charge disproportionation) and the inter-dimer FM coupling:
\begin{subequations}
  \begin{gather}
    \F=\F'\,\cos\left(\tfrac{2\phi}{3}\right)\label{EQ2.a},\\[1mm]
    S_1=\tfrac{7}{4}+\tfrac{1}{4}\,\cos\phi,\qquad 
    S_2=\tfrac{7}{4}-\tfrac{1}{4}\,\cos\phi . \label{EQ2.b}
  \end{gather}
\end{subequations}

In the limit of classical spins, the energy of the magnetic unit cell is:
\begin{multline}
  E = 4\,S_1\,S_2\Big[(\A-\Fp) + (\A-\F)\,\cos\phi + \K\,S_1\,S_2\,\cos^2\phi\Big]\\[2mm]
  + 4\,\Ad\,S_1^2\,\cos\phi - 2\D\big(S_1^2 + S_2^2\big)\big(1+\cos^2\phi\big)
\end{multline}

To assess the effect of the coupling parameters $\Ad$ and $\K$ introduced for the intermediate phase, we first evaluate the classical energy of the system varying each of them separately, and we use: $\Fp=1$ (we use $\Fp$ as energy unit, which according to fits \cite{Senff,Johnstone} typically is 10 meV), $\A=0.1$, and $\D=0.05$, as in fits to INS experiments of the laminar manganite~\cite{Senff}. 
\begin{figure}[!h]
  \centering\scalebox{0.6}{
    \includegraphics{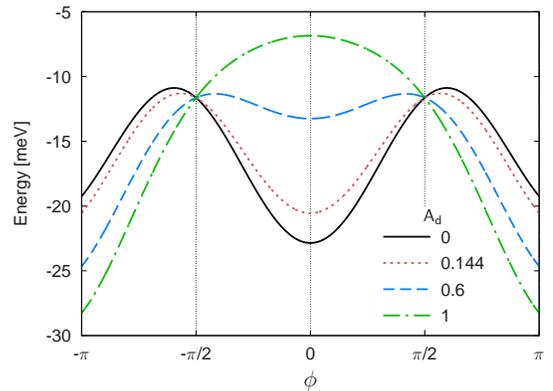}}
  \caption{(color online). Angular dependence of the classical energy of the intermediate phase, for various $\Ad$. Parameters: $\Fp=1$, $\A=0.1$, $\D=0.05$, and $\K=0$. $\F$, $S_1$ and $S_2$ as in \eqref{EQ2.a} and \eqref{EQ2.b}.}
  \label{figure2}
\end{figure} 

Fig.~\ref{figure2} shows the energy as a function of $\Ad$, when $\K=0$. As we see, only two possible classical phases are stable: $\phi=0$ or $\phi=\pi$, which are separated by the critical value $\Ad\,_{\text{crit}} =0.144$ value at which both have equal energy, and are therefore equally possible.
\begin{figure}[!h]
  \centering\scalebox{0.6}{
    \includegraphics{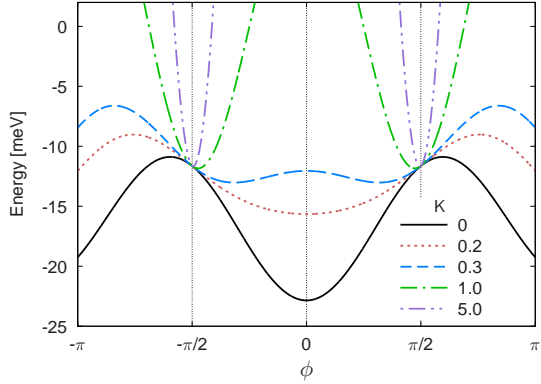}}
  \caption{(color online). Angular dependence of the energy of the classical intermediate phase, for various $\K$. Parameters: $\Ad=0$, others as in Fig.\ref{figure2}.}
  \label{figure3}
\end{figure} 

\begin{figure}[!h]
  \centering\scalebox{0.6}{
    \includegraphics{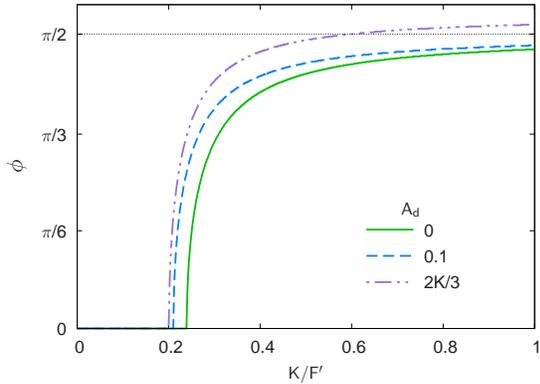}}
  \caption{(color-online). Angle characterizing the classical stable intermediate phase, as a function of $\K$ and $\Ad$. Other parameters as in Fig.\ref{figure2}.}
  \label{figure4}
\end{figure}

As shown in Fig.~\ref{figure3}, turning on $\K$ allows to shift the stable minimum of the classical energy to intermediate angles between 0 and $\pi/2$, thereby allowing the classical intermediate phase to interpolate between the classical CE and orthogonal $\pi/2$ limiting phases. Notice, however, that the $\phi=\pi/2$ state is not reached with reasonable values of $\K$ ($\sim 5\Fp$), unless we include also a finite $\Ad$ coupling, as a subtle interplay of these parameters seems to determine the stable solution. This can be appreciated in Fig.~\ref{figure4}, where we show the angle corresponding to the stable classical intermediate phase, in terms of $\K$ and $\Ad$. Recent fits to INS data in half-doped bilayered manganites ~\cite{Johnstone} provide evidence for the existence of $\Ad$. 

Notice, in any case, that an abrupt transition exists, at finite $\K$, from the stable solutions with $\phi=0$ (CE) to classically stable intermediate phase solutions, with $\phi\ne 0$. And low $\Ad$ values make it difficult to reach the intended limiting case corresponding to the orthogonal $\phi=\pi/2$ phase.

\section{Quantum treatment of the intermediate phase}
\subsection{Magnon calculation details}
To determine the magnetic excitations of the 2D quantum intermediate phase, we consider the 8-spin magnetic unit cell shown in Fig.~\ref{figure1}. We first perform a local rotation of the spin quantization axes at an angle $\phi$ for sites within the dimers rotated in the Hamiltonian~\eqref{EQ1}. The spins in the rotated system $\Sn_j^{\prime}$ are related to the fixed system $\Sn_j$ as follows: $S_j^{y\,\prime}= S_j^y$, 
\begin{subequations}
  \begin{gather}
    S_j^{x\,\prime}=S_j^x\,\cos\phi + S_j^z\,\sin\phi \, ,\notag \\
    S_j^{z\prime}=S_j^z\,\cos\phi - S_j^x\,\sin\phi \, . \notag
  \end{gather}
\end{subequations}

We then use the Holstein-Primakoff transformation for spin operators, Fourier transform, and obtain the magnon excitations by paraunitary diagonalization of the Hamiltonian matrix in the linear spin waves approximation, as in~\cite{Ventura} and \cite{Buitrago}.

\subsection{Results and discussion}
Unless otherwise stated, the following quantum magnon results, plotted along symmetry paths in the square lattice Brillouin zone, assume parameters as in the classical cases shown: $\Fp=1$, $\A=0.1$, $\D=0.05$, as consistent with recent experimental data~\cite{Senff,Ulbrich,Johnstone}, while: $\F$, $S_1$ and $S_2$ are given by \eqref{EQ2.a} and \eqref{EQ2.b}. 

\begin{figure}[!h]
  \centering\scalebox{0.6}{
    \includegraphics{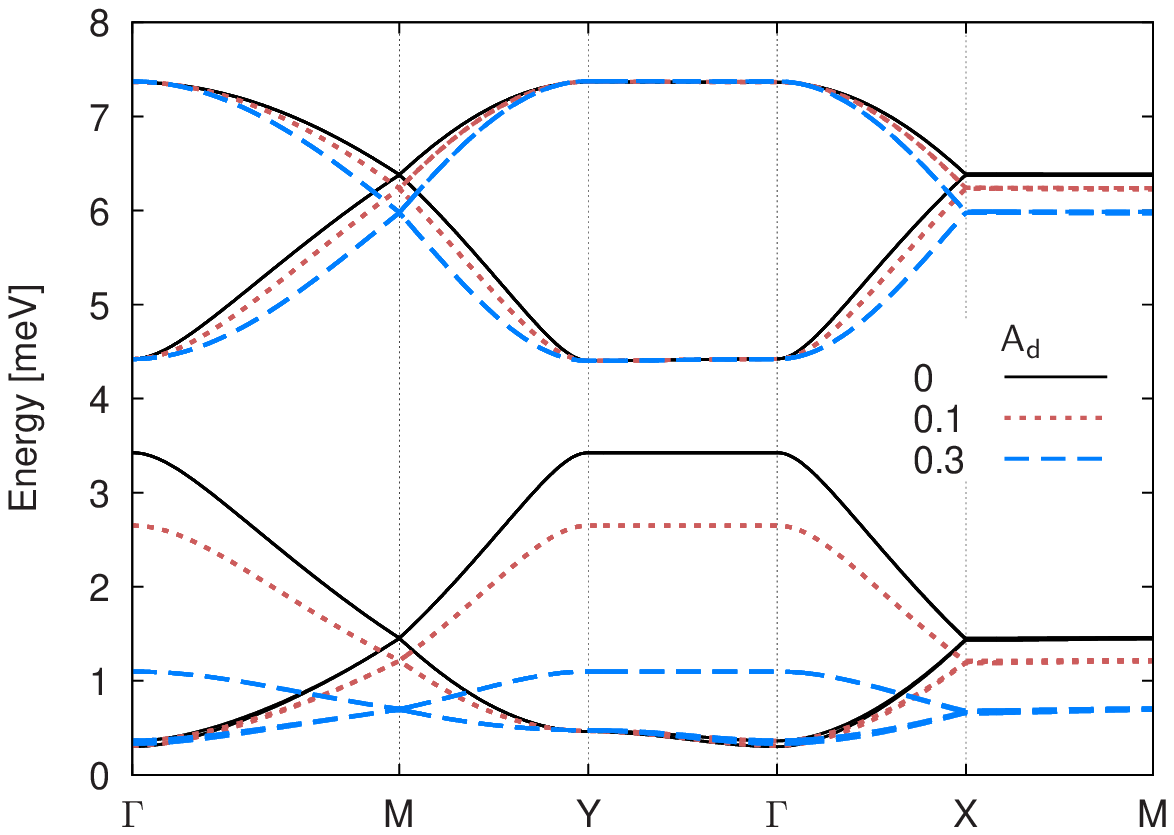}}
  \caption{(color-online). Magnons of the quantum intermediate phase, for various $\Ad$. $\K=0$, and, $\phi=15$ degrees.}
  \label{figure5}
\end{figure} 
Fig.~\ref{figure5} exhibits the effect of $\Ad$ coupling on the intermediate phase at $\phi=15$ degrees: by \eqref{EQ2.a} and \eqref{EQ2.b}, this corresponds to $\F=0.985$, $S_1=1.99$ and $S_2=1.51$. Notice that the main effect of $\Ad$ is on the lower energy branches, reducing their energy and thereby increasing the gap between the lower and higher magnon branches. 

\begin{figure}[!h]
  \centering\scalebox{0.6}{
    \includegraphics{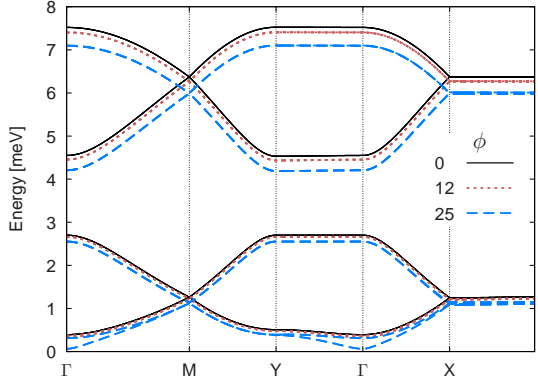}}
  \caption{(color-online). Effect of the intermediate phase angle $\phi$ on magnons. Here: $\Ad=0.1$, and $\K=0$.}
  \label{figure6}
\end{figure}

Choosing a fixed value of $\Ad=0.1$, as in Fig.~\ref{figure6} we obtain well defined magnon excitations up to an angle of 25 degrees. We see that the effect of the angle $\phi$ is to split magnon bands in some paths, due to the ensuing symmetry breaking. This effect is clearer around $\Gamma$. Although not shown here, we found that increasing $\Ad$ up to $0.43$ it is still possible to obtain well defined magnons for an intermediate phase with an angle $\phi=31$ degrees. 

\begin{figure}[!h]
  \centering\scalebox{0.6}{
    \includegraphics{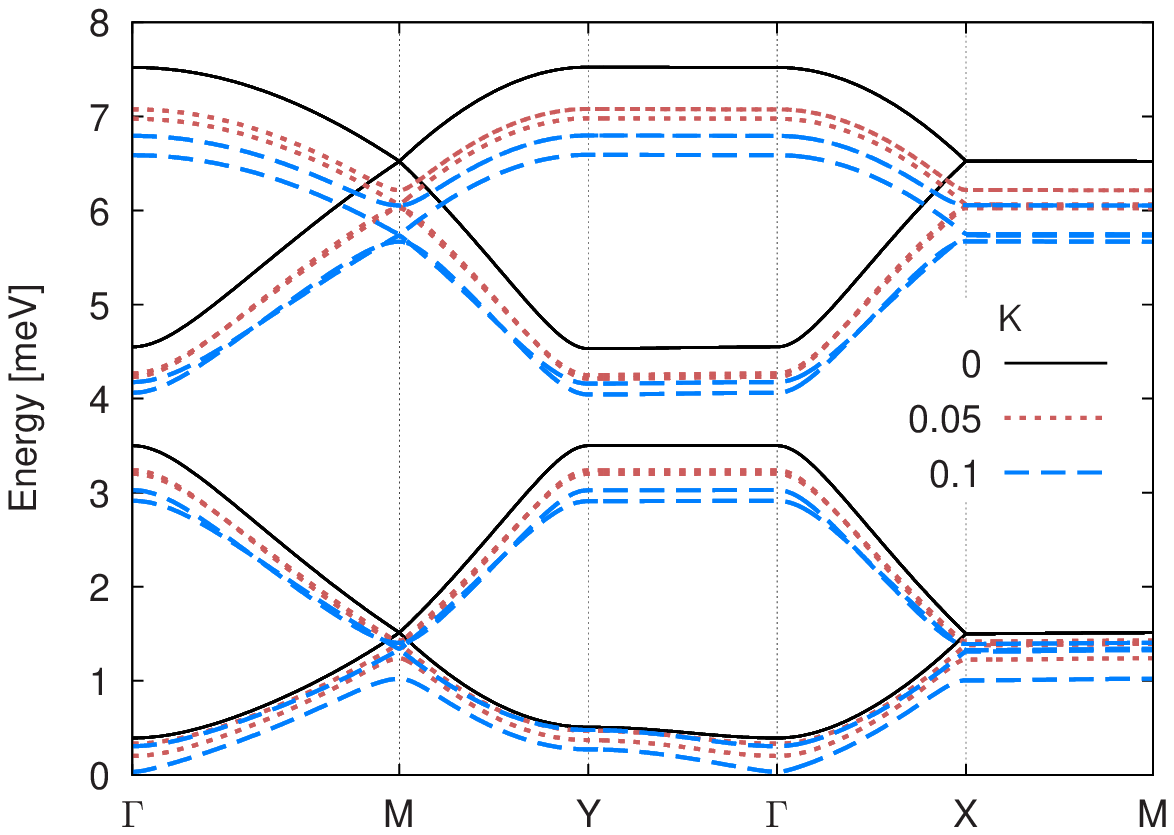}}
  \caption{(color-online).Magnons of the quantum intermediate phase, for various $\K$. \mbox{$\Ad=0$}, and $\phi=15$ degrees.}
  \label{figure7}
\end{figure}

If now $\K$ is turned on, with $\Ad=0$, we can appreciate its effect on the quantum magnons shown in Fig.~\ref{figure7}. Classically, the presence of $\K$ allowed to stabilize the intermediate phase at a wide range of angles different from $\phi=0$ and $\pi$. In the quantum case, the magnon energies decrease monotonically with an increase of $\K$. Comparing the stability of the classical and the quantum intermediate phases, we confirmed that quantum fluctuations may reduce the stability of the classical phase. For example, 
 at $\Ad=0$ the CE phase becomes unstable at $\K$ beyond the threshold value of $0.12$ while in the classical case it is stable up to $\K=0.24$, as seen in Fig.~\ref{figure4}.

As mentioned before, the subtle interplay of $\Ad$ and $\K$ introduced in our model strongly affects the stability of the intermediate phase and its excitations. We searched for pa\-ra\-me\-ter ranges leading to quantum intermediate phases stable at larger $\phi$, to see how close to the $\phi=\pi/2$ phase we could get. To extend the range of stable angles, we needed to consider larger $\Ad$, $\K$, and $\D$ values. We find that the sole consideration of longer range interactions, such as $\Ad$, is insufficient to obtain stability of the intermediate phase at a wider range of $\phi$. 

\begin{figure}[!h]
 \centering\scalebox{0.6}{
   \includegraphics{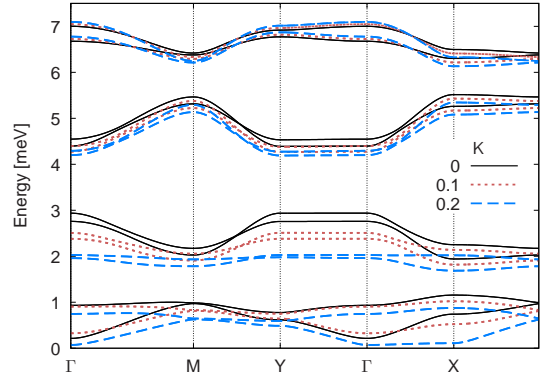}}
 \caption{(color-online). Quantum magnons of the intermediate phase at $\phi=60$ degrees, for various $\Ad$. $\K=0.5$, and $\D=0.3$.}
 \label{figure8}
\end{figure}

In Fig.~\ref{figure8} we present the magnons obtained for parameters which allow us to reach well defined spin waves for an intermediate phase at an angle of $\phi=60$ degrees. At this larger value of $\phi$, clearly more degeneracies/symmetries are broken w.r. to Fig.~\ref{figure7}. However, couplings or an anisotropy with a value considerably higher than those reported in these manganites would be required to achieve stability at higher angles, if one uses a model based purely on localized spins. 

\section{Conclusions}
We proposed a model based on interacting localized electronic spins to describe the magnetic properties of the intermediate phase of half-doped manganites, introduced by Efremov et al. in 2004~\cite{Efremov} with the idea of interpolating between the CE and the orthogonal intermediate dimer phases 
as a function of the angle between consecutive dimers along a zig-zag chain and charge disproportionation. Our model has a classically stable intermediate phase, allowing to interpolate between the desired limiting ground states. In the quantum case, we probe the stability of the intermediate phase ground state indirectly, studying the spin wave excitations resulting from it, and find a reduction of the stability of the intermediate phase. Using coupling parameters in ranges known from experimental fits of INS, we can interpolate from the CE phase up to a maximum angle of 60 degrees in the intermediate phase, not finding stable the orthogonal intermediate dimer phase proposed by Efremov et al.~\cite{Efremov}.

New INS experiments available for half-doped bilayered manganites~\cite{Johnstone} at higher energies, which have given access to the magnon higher energy branches, favour a CE-like phase with smaller charge disproportionation w.r. to \Mn{3}-\Mn{4} expected by Goodenough, and rule out the stability of the  orthogonal intermediate $\pi/2$-phase.

\section*{Acknowledgment} 
C. I. Ventura wishes to specially acknowledge discussions with D. I. Khomskii, which stimulated this work, and Prof. E. M\"uller Hartmann at the Institut f\"ur Theoretische Physik, Univ. zu K\"oln. C. I. Ventura and L. O. Manuel are Investigadores Cient\'{\i}ficos of CONICET, and acknowledge support of PICT Redes 01776 (ANPCyT) and PIP 0702 (CONICET) grants. I.~R.~Buitrago \mbox{acknowledges} support from CONICET, LAW3M organizers and PIP 0702.

\ifCLASSOPTIONcaptionsoff
 \newpage
\fi


\bibliographystyle{IEEEtran}
\bibliography{IEEEabrv,../bib/paper}
%

\end{document}